\documentclass[useAMS,usenatbib]{mn2e}
\usepackage{graphicx}

\title[Wide Field Photometry of M~22]{Wide Field Photometry of the\\ Galactic Globular Cluster
M~22\thanks{Based on observations made with the European Southern 
Observatory telescopes, using the Wide Field Imager, as part of the 
observing program 65.L-0463.}\thanks{This publication makes use
of data products from the  Two Micron All Sky Survey, which is a joint
project of the University of  Massachusetts and the Infrared Processing
and Analysis Center/California  Institute of Technology, funded by the
National Aeronautics and Space  Administration and the National Science
Foundation}}

\author[L. Monaco et al.]{L. Monaco$^{1,2}$\thanks{E-mail: 
monaco@bo.astro.it}, E. Pancino$^{1}$, F. R. Ferraro$^{2}$, M. Bellazzini$^{1}$\\
\\
$^{1}$
INAF - Osservatorio Astronomico di Bologna,  via Ranzani,1
40127 Bologna, ITALY\\ 
$^{2}$
Dipartimento di Astronomia, Universit\`a
di Bologna, via Ranzani,1 40127 Bologna, ITALY\\
}

\date{\today}

\begin{document}

\pagerange{\pageref{firstpage}--\pageref{lastpage}} \pubyear{2004}

\maketitle

\label{firstpage}

\begin{abstract}
We present wide field photometry of the Galactic Globular
Cluster M~22 in the B, V and I passbands for more than 186,000 stars.
The study is complemented by the photometry in two narrowband filters
centered on H$_{\alpha}$ and the adjacent continuum,
and by infrared J, H and K magnitudes derived from the 2~MASS survey
for $\sim$2000 stars. Profiting from this huge database, we completely
characterized the evolved stellar sequences of the cluster by determining a variety of
photometric parameters, including new photometric estimates of the mean
metallicity, reddening and distance to the cluster. In particular, from
our multi-wavelength analysis, we re-examined the long-standing
metallicity spread problem in M~22. According to our dataset, we
conclude that most of the observed width of the red giant branch must
be due to differential reddening, which amounts to a maximum of $\Delta
E(B-V)\simeq0.06$, although the presence of a small metallicity spread
cannot be completely ruled out. More specifically, the maximum
metallicity spread allowed by our data is of the order of
$\Delta$[Fe/H]$\simeq 0.1\div 0.2$ dex, i.e., not much more than what allowed
by the photometric errors. Finally, we identified most of the known 
variable stars and peculiar
objects in our field of view. In particular, we find additional
evidence supporting previous optical identifications of the central
star of the Planetary Nebula IRAS~18333-2357, which is associated with
M~22. 
\end{abstract}



\maketitle


\section{Introduction}
\label{intro}

M~22 was one of the first Galactic Globular Clusters (GGC) to be
discovered, in 1665, by Abraham Ihle and also one of the first ones to
be studied in detail \citep{shapley,arp}. It soon became the target of
a series of studies \citep[starting with][]{hesser} because of the
large color spread of its red giant branch (RGB) sequence, similar to
that observed in $\omega$~Centauri \citep{woolley}. This suggested the
possibility of a metallicity spread in M~22, as was
demonstrated in the case of $\omega$~Cen a few years before
\citep{dickens}.

However, while the presence of significant reddening variations was excluded in the
case of $\omega$~Cen \citep{cannon80}, some differential reddening was
found in the direction of M~22 \citep[see][ and references therein]{rich}.
Of course, the presence of differential reddening does not exclude the
presence of some metallicity spread, since the two effects could be
both present and responsible for the observed width of the RGB. The
photometric studies of M~22 by \citet{zoc} and \citet{rich} were able
to put upper limits to the amount of differential reddening of 
$\sigma_{\Delta E(V-I)}$=0.05 and $\Delta$E(B$-$V)=0.07, respectively.
Interestingly, \citet{rich} demonstrated that part of the
spread observed in the Str\"omgren colours must be due to CH and CN 
variations, so that the eventual spread in heavy elements should be negligible.

Spectroscopic studies, on the other hand, gave controversial results as
far as the heavy elements abundances are concerned, while a spread in
the CH and CN abundances of RGB stars appears unquestionable
\citep{nofre83}\footnote{Some marginal evidence for an over-abundance
of $s$-process elements has also been reported by \citet{gratton}.}. For
example, some studies reported on metallicity variations of
0.3$\div$0.5 dex in Ca and/or Fe, often correlated with the CH and CN
variations \citep{ruth,cathy,leh,brown}, while other studies found
no significant variation in the heavy element content
\citep{manduca,cohen,gratton,go89,laird}. 

Concerning this apparent contradiction, \citet{leh} noted that, since
the the estimated standard deviation of the abundance variations
($\sim$0.2 dex) is close to the typical uncertainty of most
spectroscopic analyses, it is very difficult to unequivocally
demonstrate the presence of a metallicity spread. This is particularly
true if one considers that most of the above studies are only based on
a handful of stars ($\leq 10$). In the next years, thanks to the new
generation of multi-object spectrographs, it will be possible to analyze
large samples of stars in a homogeneous way, thus shedding more light
on this issue.

M~22 is a metal poor ([Fe/H]$\simeq$-1.62, \citet{harris}) 
and very bright cluster. Considering also its position on the sky, (l; b)=(9.89; -7.55), 
and its proximity to us (only 3.2~Kpc from the Sun\footnote{Note that the Galactic Bulge is 
in the background of M~22.}) M~22 is certainly 
the ideal target for various studies, ranging from the dynamical 
modelling of dense stellar system \citep{dina} to microlensing studies 
\citep{sahu}. On the other hand
the characterization of its stellar content is still quite uncertain 
since it suffers, once again, from the presence of differential reddening 
along the line of sight. 
Here we provide a complete and homogeneous photometric
characterization of the stellar content of M~22. 

The paper is organized as follows. In Section~\ref{sec-obs} we present
the observations and data reduction procedures, compare our results
with previous literature and derive optical and infrared mean ridge
lines (MRL). In Section~\ref{sec-red} we deal with the differential
reddening and metallicity spread issues. In Section~\ref{sec-par} we
derive the mean metallicity, reddening and distance along with other photometric
parameters of M~22. In Section~\ref{sec-var} we identify the known
variables and peculiar objects, including the central star of the Planetary Nebula 
IRAS~18333-2357. In Section~\ref{sec-con} we summarize our main
results.


\section{Observations and Data Reduction}
\label{sec-obs}

Observations were obtained at the 2.2m ESO/MPI telescope at la Silla,
Chile, using the {\em Wide Field  Imager} (WFI), a mosaic of eight
$2048\times4096$ pixels CCDs. The instrument scale is
$0\farcs 238$~pix$^{-1}$, giving a total field of view of $34' \times 33'$.
A set of B, V and I images were secured during a single observing run
on 6-7 July 2000, with exposure times ranging from 5 to 400 sec. We
also secured a set of exposures in two narrowband filters, H$_{\alpha}$
and $\Re$, centered around the H$_{\alpha}$ line
($\lambda_c\simeq$6580~\AA) and on the adjacent continuum
($\lambda_c\simeq$6650~\AA). The un-calibrated H$_{\alpha}$ photometry
is briefly discussed in Section~\ref{sec-pn}. The average seeing
during the observations was $\sim$1$\arcsec$ full width at half maximum (FWHM). 

The raw images were corrected for bias and flat-field using specific 
IRAF\footnote{IRAF is distributed by the National Optical Astronomy 
Observatories, which is operated by the association of Universities for
Research in Astronomy, Inc., under contract with the National Science
Foundation.} procedures, within the {\em noao.mscred} package. The
photometric reduction was carried out using the DAOPhot~II and ALLSTAR
packages \citep{daophot}. Stars were searched independently on each 
CCD of the WFI mosaic with a 3\,$\sigma$ threshold and fitted with a 
second order spatially variable point spread function (PSF). 

We used standard IRAF routines to obtain aperture photometry 
for a sample of isolated stars at various positions along the CCD. 
We derived the optimal radius for the aperture photometry by constructing the 
curve of growth of each star. We compared the aperture photometry with
the DAOPHOT PSF-fitting photometry and we obtained an aperture correction of 
0.00 for the V and I filters (with a typical error of 0.02 and 0.01 
respectively) and 0.03 for the B filter (with a typical error of 0.02). 
The aperture corrections do not correlate with the position of the star on
the CCD.

Then we corrected our instrumental magnitudes considering 
the extinction coefficients available for each filter from the ESO web 
page\footnote{http://www.eso.org/gen-fac/pubs/astclim/lasilla/index.html,\\ 
see also: http://www.ls.eso.org/lasilla/sciops/2p2/E2p2M/WFI/\\zeropoints/} 
and the airmass at the beginning of the observations.
 
All observations were carried out under photometric conditions. The
calibration to the standard Johnson-Cousins photometric system was
obtained  using two standard fields \citep[namely TPhe and PG~1323,
][]{landolt} observed at different airmasses during the night. The
adopted calibrating equations are:
\begin{eqnarray*}
\rm B&=&b+0.45~(b-v)-0.48\\
\rm V&=&v-0.09~(b-v)-1.03\\
\rm I&=&i+0.12~(v-i)-1.88
\end{eqnarray*}    
where b, v and i are the corrected instrumental magnitudes and B, V and I the
corresponding magnitudes in the Johnson-Cousins photometric system.

\begin{figure} 
\begin{center}
\hspace{-2.6cm}
\includegraphics[width=180mm,height=220mm,angle= 0]{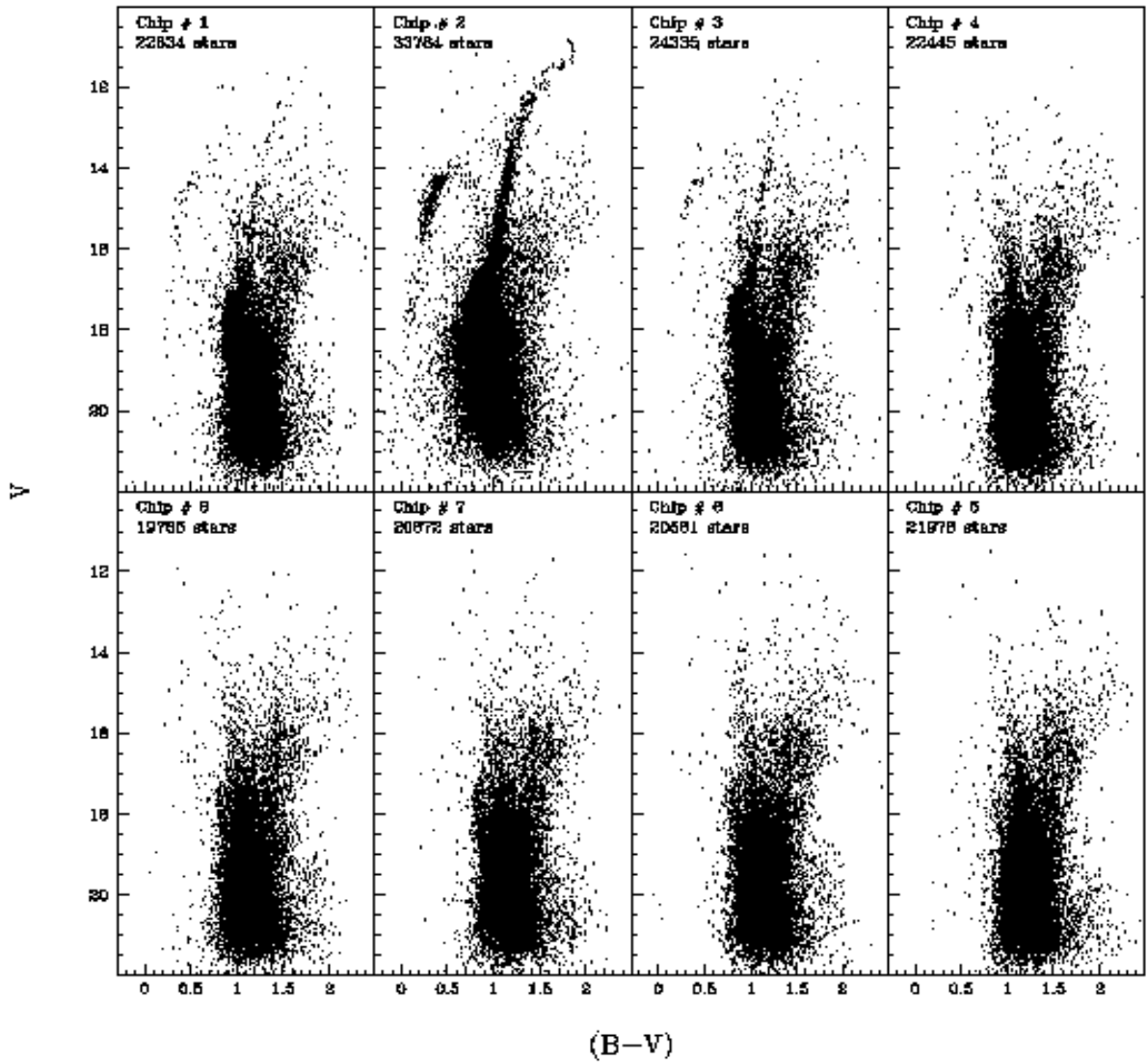}

\caption{The V {\it vs} (B-V) CMDs obtained for each of the eight 
chips of the ESO Wide Field Imager mosaic.} 
\label{f1} 
\end{center}
\end{figure}

\begin{figure} 
\includegraphics[width=180mm,height=220mm]{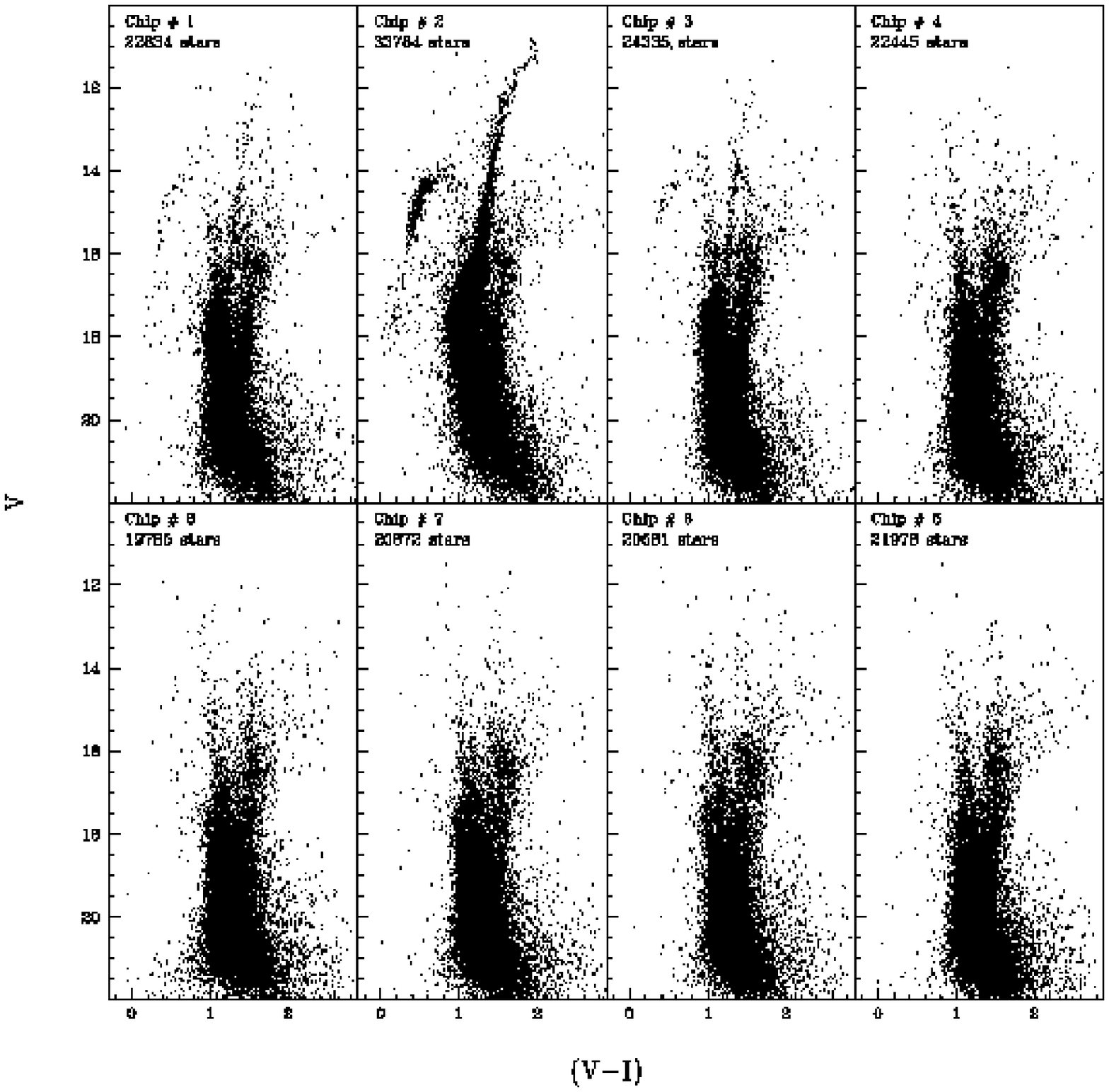}

\caption{The V {\it vs} (V-I) CMDs obtained for each of the eight chips 
of the ESO Wide Field Imager mosaic.} 

\label{f1bis} 
\end{figure}

The resulting, calibrated color magnitude diagrams (CMDs) are displayed
on Figures ~\ref{f1} and \ref{f1bis} in the V,(B$-$V) and V,(V$-$I) planes,
respectively. As can be seen, the population of M~22 dominates the CMD
resulting from CCD \#2, since the cluster center has been placed on
that chip. The bulge and disk populations dominate instead the CMDs of
the outer CCDs, where the contribution by M~22 tends to disappear (see,
e.g., chips \#4 and \#5). A few bright stars with V$<$11.3, (B$-$V)=1.8
and (V$-$I)=1.9 are saturated. 

Therefore, in the following sections, we will restrict our analysis to
stars measured in CCD \#2 only, since most of the cluster population
lies in that chip, where the contamination by the disk and the bulge is
less important and the CMD 
\clearpage
\clearpage
\clearpage
\begin{figure} 
\includegraphics[width=84mm]{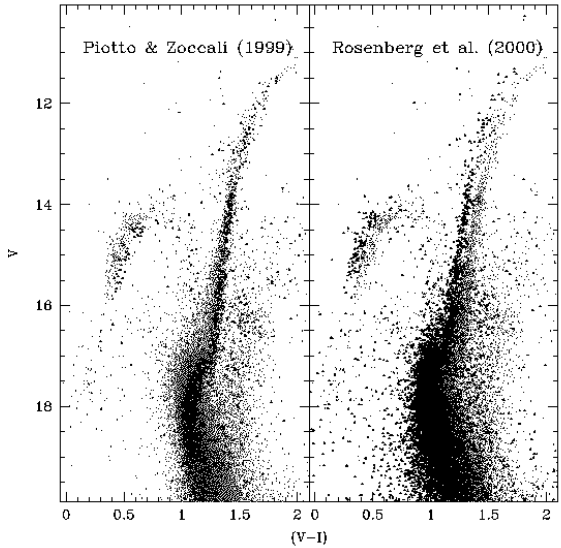}

\caption{The M~22 photometry 
by \citet{zoc} (left panel) and by \citet{rosen} (right panel) is superposed 
to the CMD obtained in this paper (light grey).} 

\label{confr} 
\end{figure} 

\begin{figure} 
\includegraphics[width=84mm]{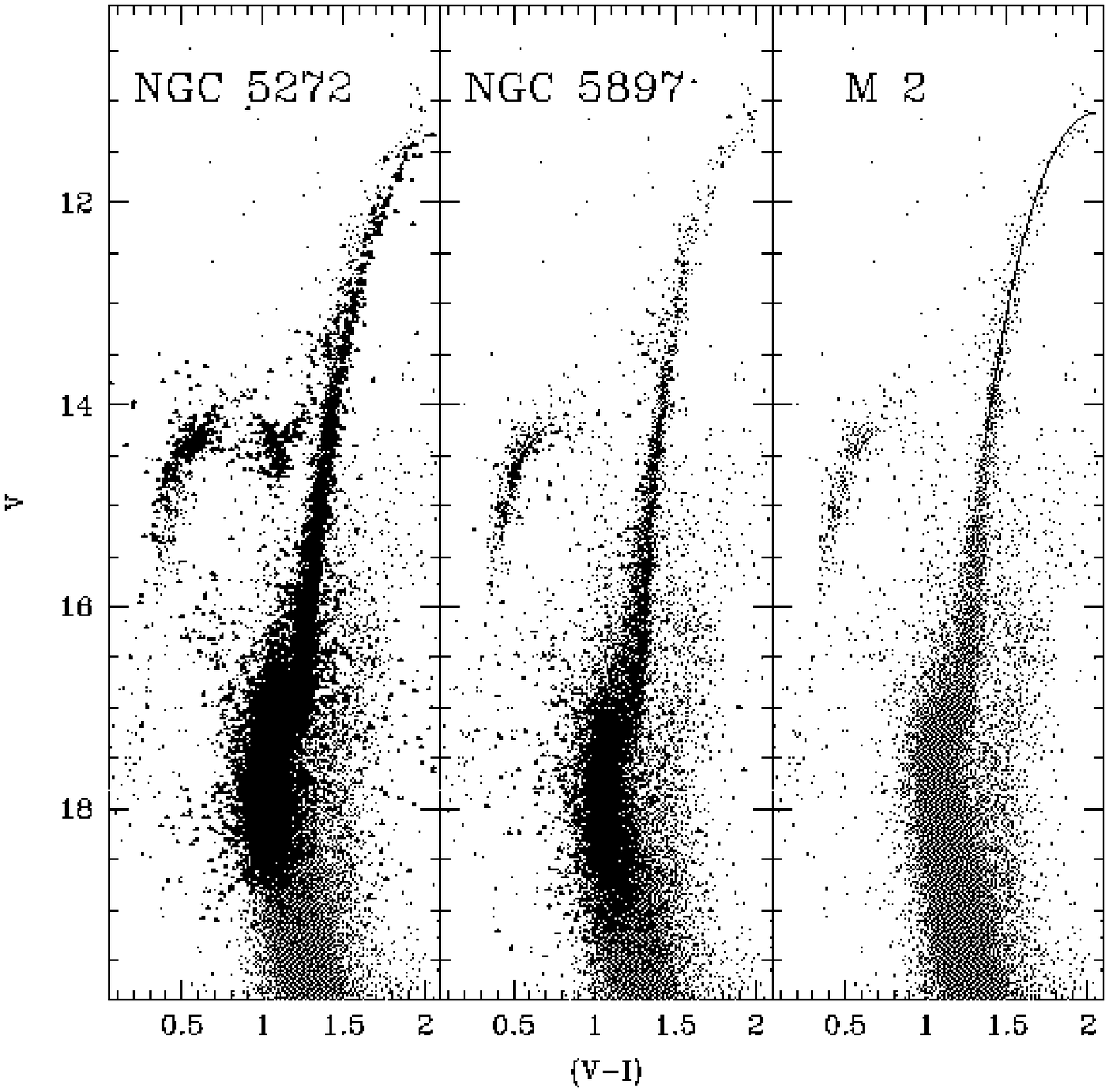}

\caption{The photometries of NGC~5272 (left panel), NGC~5897 (middle panel) and the 
mean RGB ridge line of M~2 
(right panel) are superposed on our M~22 CMD (light grey). See the text for references.} 

\label{confr2} 
\end{figure} 
\begin{figure} 
\includegraphics[width=84mm]{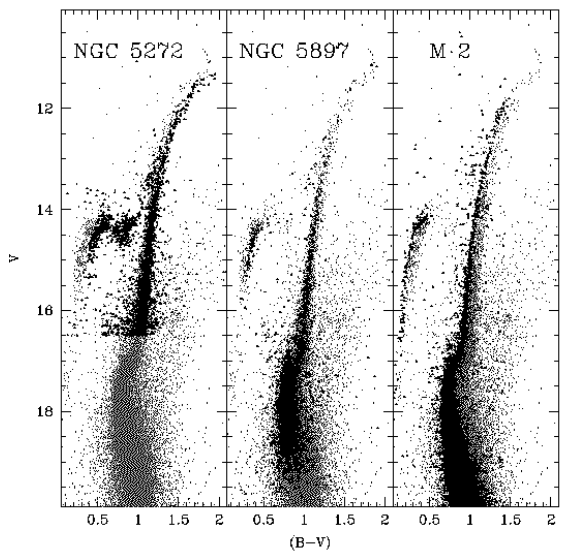}

\caption{The photometries of NGC~5272 (left panel), NGC~5897 (middle panel) 
and  M~2 (right panel) are superposed on our M~22 CMD (light grey). 
See the text for references.} 

\label{confr3} 
\end{figure} 
\hspace{-0.7cm}appears cleaner. This way, we also avoid the
propagation of the photometric zero-point differences between~different~CCDs, 
which could degrade the overall quality of the
photometric catalog. 

Finally, in order to complement the multi-wavelenght study of M~22, we
identified approximately 1840 stars in our catalogue having J, H and
K$_s$ magnitudes measured by the 2~MASS survey\footnote{http://www.ipac.caltech.edu/2mass}. 
An example of the
resulting K$_S$,(V$-$K$_S$) and K$_S$,(J$-$K$_S$) CMDs can be seen in Fig.~\ref{2mass}.

%

\subsection{Online Catalogue}
Although the following analysis is entirely based on stars belonging to
CCD \#2, several M~22 stars are still present in CCDs \#1, \#3, \#6, \#7 and \#8
(Figures~\ref{f1} and \ref{f1bis}). Therefore, the full catalog presented
in Table~\ref{tabcat} and published electronically contains all the stars detected in these 
CCDs. 

The catalog contains the B, V, I calibrated magnitudes on columns \#2, \#4, \#6, 
each followed by the formal daophot errors ($\delta\rm$B, $\delta\rm$V, 
$\delta\rm$I). Only stars measured both in the V and I filters are tabulated 
while the B magnitude is provided only when available. A flag ``0.000" can 
be found on columns \#2 and \#3 if a star is not measured in the B filter.   

Star positions are provided both in the pixel coordinate system of each CCD (columns
\#7, \#8) and in the equatorial (RA; Dec) coordinate system (columns
\#9, \#10). In the first column we provide a sequential identifier for each star.

The coordinates in the J2000.0 absolute astrometric system 
have been obtained with a procedure
already described in other papers \citep[see e.g.][]{f01}. The new
astrometric {\it Guide Star Catalog} (GSC~II) recently released and now
available on the web\footnote{{\rm
http://www-gsss.stsci.edu/gsc/gsc2/GSC2home.htm}}, was used as
reference. More than two thousand GSC~II astrometric reference stars have been
found in the field of view of each chip, 
allowing for an accurate absolute positioning of the image. In
order to derive an astrometric solution for each WFI CCD, 
we used a program specifically developed at the Bologna Observatory (P.
Montegriffo et al 2004, in preparation). As a result of the entire
procedure, r.m.s. residuals of  $\sim 0\farcs 2$, 
both in RA and Dec, were obtained. This
value can be considered as a representative uncertainty of the
astrometric calibration procedure. 

The photometry of stars belonging to CCDs \#1, \#3, \#6, \#7 
and \#8 has been shifted to match the photometry of CCD \#2, 
taking into account the different CCD responses. 
The correction was calculated by fitting the mean ridge lines
calculated in \S\ref{sec-mrl} to the CMDs.  
The correction applied to each of the B, V, I magnitudes of the external chips is 
always $\leq$0.1 mag. 
However, in the CMD of chips \#6, \#7 and \#8, only a handful of cluster
stars is present in the main sequence and turn off regions and the 
calculated corrections are correspondingly less certain.

\begin{table*}

\caption[]{CCD B, V, I online photometric catalog of M~22. Only a few entries are displayed to illustrate 
the catalog format and contents.}

\label{tabcat}
$$
\begin{array}{ccccccccccc}
\hline 
&&&&&& Chip \#2 &&\\
\hline \hline
\rm{Star~Id} & \rm B & \delta \rm B & \rm V & \delta \rm V & \rm I 
& \delta \rm I & X_{pix} & Y_{pix} &
\rm{RA} & \rm{Dec}\\
\hline
1& 12.071& 0.010&10.351& 0.010& 8.544&0.124&  806.273& 1876.142& 279.11637069&-23.92032070\\
2& 10.854& 0.010&10.483& 0.011& 9.938&0.014& 1808.769& 4086.802& 279.04357491&-23.77422799\\
3& 12.632& 0.010&10.835& 0.010& 8.925&0.015& 1221.985& 1819.724& 279.08629030&-23.92412172\\
4& 12.707& 0.010&10.885& 0.010& 8.936&0.018& 1806.331& 1933.347& 279.04400418&-23.91672994\\
5& 12.793& 0.010&10.941& 0.010& 8.979&0.015&   52.285& 1092.889& 279.17106212&-23.97201436\\
6& 12.888& 0.010&11.043& 0.010& 9.123&0.015& 1406.258& 2075.037& 279.07292590&-23.90726310\\
7& 12.958& 0.010&11.095& 0.010& 9.096&0.018&  333.186&  757.506& 279.15076070&-23.99417931\\
\hline
\end{array}
$$
\end{table*}


\subsection{Literature Comparisons}\label{confronti}

In order to check our calibration, we compared our V,(V$-$I) photometry
with two catalogues previously published by \citet{zoc} and
\citet{rosen}.

Fig.~\ref{confr} shows the results of such a comparison. In particular,
the photometry by \citet{zoc} appears in good agreement with ours as
far as the upper RGB is concerned (left panel of Fig.~\ref{confr}).
However, their horizontal branch (HB) appears {\em redder} and {\em
fainter} than ours, by approximately 0.18~dex in magnitude and
0.04~dex in color.  A similar discrepancy can be observed for
the lower RGB and the turn-off (TO) regions, pointing towards a
possible residual (V$-$I) color-term between the two calibrations. On
the other hand, the comparison with \citet{rosen} results again in a
discrepancy, but in the opposite sense (right panel of
Fig.~\ref{confr}). Both the RGB and HB appear in fact systematically
{\em bluer} and {\em brighter} than ours, by roughly 0.1~dex in
color and 0.05~dex in magnitude. In this case, there appears to be a
residual zero-point difference between the two calibrations. 

To further investigate on this issue, we compared (Fig.~\ref{confr2})
our photometry with that of three globular clusters having similar
metallicities (and thus similar RGB shapes) as M~22. More
specifically, the left panel of Fig.~\ref{confr2} shows the comparison
of our M~22 photometry with that of NGC~5272 \citep{f2}, which has a
metallicity of [Fe/H]$_{ZW}$=-1.66 in the \citet{zw} scale. NGC~5272
has been corrected with the reddening and distance tabulated by
\citet{f99} and with the corresponding values for M~22 
(see Section~\ref{sec-par}). The middle panel of
Fig.~\ref{confr2} shows the comparison with NGC~5897 \citep{f1}, which
has [Fe/H]$_{ZW}$=-1.68 and has been corrected for reddening and
distance as above. Finally, the right panel of Fig.~\ref{confr2} shows
the comparison of our M~22 photometry with the RGB mean ridge line of
M~2, which has  [Fe/H]$_{ZW}=-1.62$ \citep{harris} and was published by
\citet{m2} already corrected for reddening and distance. We thus
applied the distance modulus and reddening of M~22 following
\citet{harris}, to be consistent with the distance scale of
\citet{m2}. 

As can be seen, all the three panels of Fig.~\ref{confr2} show an
excellent match with the present photometry of M~22, thus dispelling
any remaining doubt on the adopted absolute V and I calibration. 

Unfortunately, among the recent CCD studies of M~22, the
only available B-band dataset is the one by \citet{kalu} and, as the
authors explicitely state in their paper, their {\em absolute}
photometric calibration is not reliable. 
In Fig.~\ref{confr3} we compared our photometry to that of NGC~5272, NGC~5897 
\citep{f99} and M~2 \citep{m2bv} in the V {\it vs} B-V plane. The CMDs of 
NGC~5272 and NGC~5897 have been corrected exactly as in Fig.~\ref{confr2}. 
The M~2 CMD have been corrected for the appropriate reddening and distance 
and the corresponding M~22 values tabulated by \citet{harris}. 
We find a reasonable agreement with each of the reference clusters and we 
conclude that also our B-band calibration can be considered reliable. 
In particular, a good match is obtained in the case of 
NGC~5897 and of the RGB of NGC~5272. The NGC~5272 HB is redder than the 
M~22 one even if the mean level appears similar. We also obtain a reasonable
agreement in the case of M~2, even if its CMD is somewhat bluer than the M~22
one.


\subsection{Optical and IR Mean Ridge Lines}
\label{sec-mrl}

\begin{figure} 
\includegraphics[width=84mm]{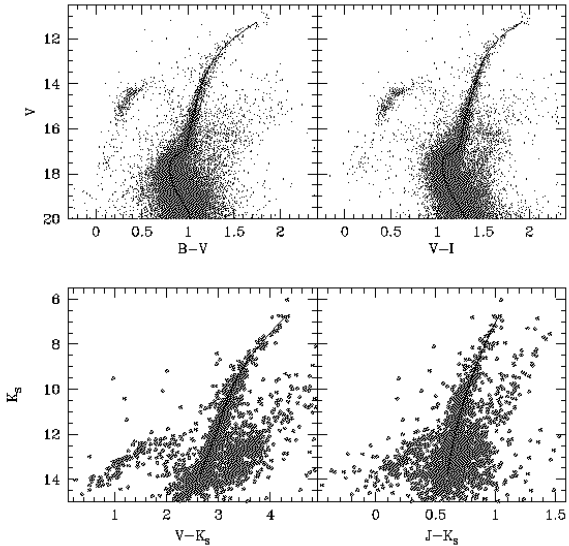}

\caption{The mean ridge lines listed in Table \ref{lm} and \ref{lmk} are superposed 
to the CMDs in the optical (upper panels), infrared (lower-right panel) and 
optical-infrared planes (lower-left panel).} 

\label{2mass} 
\end{figure}

Profiting from this large photometric database, we have constructed the
RGB mean ridge lines for M~22, both in the optical and infrared colors.
Since the optical photometry is deeper than the infrared 2~MASS
photometry, we were able to reach down to the main sequence in the
V,(V$-$I) and V,(B$-$V) planes, while we reach the base of the
RGB in the K,(V$-$K$_S$) and K,(J$-$K$_S$) planes (see Fig.~\ref{2mass}). 

To derive the lines, we sliced the sequences in appropriate magnitude
and/or color bins of variable size (depending on the number of points
and on the shape of the sequence). The representative point of each bin has
been computed as the 2\,$\sigma$ clipped average. The set of
representative points has then been fitted by analytical polynomials
of variable degree until the best-fitting polynomial was found. The
resulting mean ridge lines are tabulated in Tables~\ref{lm} (optical)
and \ref{lmk} (optical-infrared) and they are overplotted on
the respective CMDs, in Fig.~\ref{2mass}.

\begin{figure} 
\includegraphics[width=84mm]{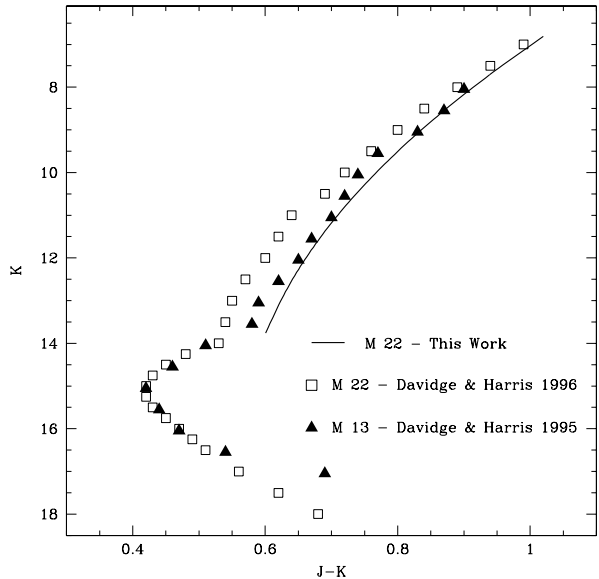}

\caption{Our M~22 infrared mean ridge line (continuous line) is compared 
with the M~22 and M~13 normal points by \citet{davi1} (open squares) and 
\citet{davi2} (filled triangles), respectively. We applied horizontal and
vertical shifts to align M~13 to the M~22 normal points at the turn off
and the point 0.05 mag redward of the turn off, just as done by 
\citet{davi1} (see lower panel in their figure 7).} 

\label{davi} 
\end{figure}

The morphology of the M~22 RGB in the infrared passbands was studied by 
\citet{davi1}. These authors noted a pronounced discrepancy in the lower RGB 
between their M~22 (K, J-K) fiducial line and that of M~13 \citep[][]{davi2}, a cluster 
of similar metallicity \citep{harris}.
We compared the shape of our (K$_S$, J-K$_S$) fiducial line 
\citep[converted to the standard system using the relation of][]{carp} to 
the one of M~13 published by \citet[][]{davi2} and we find a reasonable
agreement among the two ridge lines (see figure \ref{davi}).


\section{Differential Reddening and Metallicity Spread} 
\label{sec-red}

\begin{table*}

\caption[]{Optical M~22 mean ridge lines}

\label{lm}
$$
\begin{array}{ccc|ccc|ccc|ccc|ccc}
\hline \hline		 
$V$ & $(B--V)$ & $(V--I)$ & $V$ & $(B--V)$ & $(V--I)$ &
$V$ & $(B--V)$ & $(V--I)$ & $V$ & $(B--V)$ & $(V--I)$ & 
$V$ & $(B--V)$ & $(V--I)$  \\					 
\hline			 
11.25 & 1.752 & 1.928 & 13.49 & 1.203 & 1.452 & 15.68 & 1.040 & 1.298 & 17.47 & 0.812 & 1.060 &  18.57 & 0.845 & 1.117 \\    
11.38 & 1.703 & 1.877 & 13.60 & 1.190 & 1.444 & 15.78 & 1.034 & 1.293 & 17.52 & 0.809 & 1.059 &  18.62 & 0.851 & 1.122 \\    
11.49 & 1.659 & 1.830 & 13.69 & 1.181 & 1.436 & 15.87 & 1.029 & 1.287 & 17.57 & 0.805 & 1.058 &  18.67 & 0.857 & 1.127 \\       
11.60 & 1.621 & 1.796 & 13.78 & 1.171 & 1.429 & 15.97 & 1.024 & 1.283 & 17.62 & 0.803 & 1.057 &  18.72 & 0.862 & 1.133 \\      
11.69 & 1.593 & 1.763 & 13.89 & 1.159 & 1.422 & 16.07 & 1.019 & 1.278 & 17.67 & 0.801 & 1.057 &  18.78 & 0.870 & 1.140 \\      
11.79 & 1.561 & 1.733 & 13.99 & 1.149 & 1.415 & 16.18 & 1.013 & 1.273 & 17.73 & 0.800 & 1.057 &  18.82 & 0.874 & 1.144 \\      
11.89 & 1.528 & 1.705 & 14.08 & 1.141 & 1.409 & 16.27 & 1.008 & 1.269 & 17.77 & 0.799 & 1.058 &  18.86 & 0.880 & 1.149 \\      
11.99 & 1.496 & 1.679 & 14.18 & 1.135 & 1.402 & 16.37 & 1.002 & 1.264 & 17.82 & 0.800 & 1.059 &  18.92 & 0.888 & 1.156 \\    
12.09 & 1.470 & 1.654 & 14.27 & 1.128 & 1.395 & 16.47 & 0.997 & 1.259 & 17.87 & 0.800 & 1.061 &  18.99 & 0.897 & 1.164 \\    
12.18 & 1.446 & 1.632 & 14.38 & 1.118 & 1.388 & 16.57 & 0.993 & 1.255 & 17.92 & 0.802 & 1.063 &  19.03 & 0.903 & 1.169 \\    
12.27 & 1.423 & 1.610 & 14.48 & 1.111 & 1.380 & 16.68 & 0.988 & 1.246 & 17.97 & 0.803 & 1.065 &  19.07 & 0.909 & 1.175 \\    
12.38 & 1.397 & 1.591 & 14.57 & 1.105 & 1.373 & 16.77 & 0.982 & 1.236 & 18.02 & 0.805 & 1.068 &  19.11 & 0.915 & 1.180 \\    
12.47 & 1.374 & 1.573 & 14.67 & 1.098 & 1.364 & 16.88 & 0.969 & 1.224 & 18.07 & 0.809 & 1.072 &  19.21 & 0.929 & 1.194 \\    
12.57 & 1.354 & 1.557 & 14.78 & 1.093 & 1.356 & 16.97 & 0.950 & 1.211 & 18.12 & 0.811 & 1.076 &  19.31 & 0.944 & 1.209 \\    
12.67 & 1.333 & 1.541 & 14.87 & 1.087 & 1.350 & 17.08 & 0.912 & 1.171 & 18.17 & 0.814 & 1.080 &  19.41 & 0.959 & 1.225 \\    
12.76 & 1.315 & 1.527 & 14.97 & 1.080 & 1.342 & 17.14 & 0.885 & 1.111 & 18.22 & 0.817 & 1.085 &  19.50 & 0.971 & 1.237 \\    
12.86 & 1.297 & 1.515 & 15.08 & 1.074 & 1.335 & 17.17 & 0.875 & 1.098 & 18.27 & 0.820 & 1.090 &  19.60 & 0.985 & 1.255 \\    
12.95 & 1.281 & 1.503 & 15.17 & 1.069 & 1.329 & 17.22 & 0.860 & 1.082 & 18.32 & 0.823 & 1.094 &  19.70 & 0.998 & 1.273 \\    
13.06 & 1.263 & 1.492 & 15.27 & 1.063 & 1.323 & 17.27 & 0.845 & 1.075 & 18.37 & 0.827 & 1.098 &  19.80 & 1.011 & 1.291 \\    
13.16 & 1.249 & 1.482 & 15.38 & 1.056 & 1.316 & 17.32 & 0.833 & 1.070 & 18.42 & 0.831 & 1.103 &  19.91 & 1.023 & 1.310 \\    
13.28 & 1.230 & 1.470 & 15.49 & 1.050 & 1.311 & 17.37 & 0.824 & 1.065 & 18.47 & 0.835 & 1.108 &  20.01 & 1.031 & 1.330 \\    
13.40 & 1.214 & 1.460 & 15.58 & 1.045 & 1.305 & 17.42 & 0.817 & 1.062 & 18.52 & 0.839 & 1.112 &  && \\  			 
\hline			     
\end{array}		     
$$			     
\end{table*}		     
			     
\begin{table*}		     
			     
\caption[]{Infrared and Optical-Infrared M~22 mean ridge lines}
			     
\label{lmk}		     
$$			     
\begin{array}{ccc|ccc|ccc|ccc|ccc}
\hline \hline		     
$K$_S & $(V--K$_S$)$ & $(J--K$_S$)$ & $K$_S & $(V--K$_S$)$ & $(J--K$_S$)$ &
$K$_S & $(V--K$_S$)$ & $(J--K$_S$)$ & $K$_S & $(V--K$_S$)$ & $(J--K$_S$)$ & 
$K$_S & $(V--K$_S$)$ & $(J--K$_S$)$ \\					 
\hline			     
6.814	 & 4.269   & 1.020 & 8.234    & 3.723	& 0.895 &  9.653    & 3.319   & 0.790 &  11.07    & 3.075   & 0.706 & 12.49    & 2.863   & 0.641\\	      
6.888	 & 4.240   & 1.013 & 8.308    & 3.698	& 0.889 &  9.728    & 3.303   & 0.785 &  11.15    & 3.064   & 0.702 & 12.57    & 2.850   & 0.639\\	      
6.963	 & 4.210   & 1.006 & 8.383    & 3.672	& 0.883 &  9.803    & 3.287   & 0.780 &  11.22    & 3.054   & 0.698 & 12.64    & 2.837   & 0.636\\	      
7.038	 & 4.180   & 0.999 & 8.458    & 3.647	& 0.877 &  9.878    & 3.272   & 0.775 &  11.30    & 3.043   & 0.694 & 12.72    & 2.824   & 0.633\\	      
7.113	 & 4.150   & 0.992 & 8.533    & 3.623	& 0.871 &  9.952    & 3.257   & 0.771 &  11.37    & 3.033   & 0.690 & 12.79    & 2.811   & 0.631\\    	 	  
7.187	 & 4.120   & 0.985 & 8.607    & 3.599	& 0.865 &  10.03    & 3.243   & 0.766 &  11.45    & 3.023   & 0.687 & 12.87    & 2.797   & 0.628\\    	 	  
7.262	 & 4.090   & 0.979 & 8.682    & 3.576	& 0.860 &  10.10    & 3.229   & 0.761 &  11.52    & 3.012   & 0.683 & 12.94    & 2.784   & 0.626\\	 	  
7.337	 & 4.061   & 0.972 & 8.757    & 3.553	& 0.854 &  10.18    & 3.215   & 0.757 &  11.60    & 3.002   & 0.680 & 13.02    & 2.771   & 0.623\\	 	  
7.412	 & 4.031   & 0.965 & 8.831    & 3.531	& 0.848 &  10.25    & 3.202   & 0.752 &  11.67    & 2.991   & 0.676 & 13.09    & 2.757   & 0.621\\	 	  
7.486	 & 4.002   & 0.958 & 8.906    & 3.509	& 0.843 &  10.33    & 3.189   & 0.748 &  11.75    & 2.980   & 0.673 & 13.17    & 2.744   & 0.618\\	 	  
7.561	 & 3.973   & 0.952 & 8.981    & 3.487	& 0.837 &  10.40    & 3.176   & 0.743 &  11.82    & 2.969   & 0.669 & 13.24    & 2.731   & 0.616\\	 	  
7.636	 & 3.944   & 0.945 & 9.056    & 3.467	& 0.832 &  10.48    & 3.164   & 0.739 &  11.90    & 2.958   & 0.666 & 13.32    & 2.718   & 0.614\\	 	  
7.710	 & 3.915   & 0.939 & 9.130    & 3.446	& 0.826 &  10.55    & 3.152   & 0.734 &  11.97    & 2.947   & 0.663 & 13.39    & 2.706   & 0.612\\	 	  
7.785	 & 3.887   & 0.932 & 9.205    & 3.427	& 0.821 &  10.62    & 3.141   & 0.730 &  12.04    & 2.936   & 0.660 & 13.46    & 2.694   & 0.610\\	 	  
7.860	 & 3.858   & 0.926 & 9.280    & 3.407	& 0.816 &  10.70    & 3.129   & 0.726 &  12.12    & 2.924   & 0.656 & 13.54    & 2.682   & 0.607\\	 	  
7.935	 & 3.831   & 0.920 & 9.355    & 3.389	& 0.810 &  10.77    & 3.118   & 0.722 &  12.19    & 2.912   & 0.653 & 13.61    & 2.671   & 0.605\\	 	  
8.009	 & 3.803   & 0.913 & 9.429    & 3.370	& 0.805 &  10.85    & 3.107   & 0.718 &  12.27    & 2.900   & 0.650 & 13.69    & 2.662   & 0.603\\	 	  
8.084	 & 3.776   & 0.907 & 9.504    & 3.353	& 0.800 &  10.92    & 3.096   & 0.714 &  12.34    & 2.888   & 0.647 & 13.76    & 2.653   & 0.601\\	 	  
8.159	 & 3.750   & 0.901 & 9.579    & 3.336	& 0.795 &  11.00    & 3.085   & 0.710 &  12.42    & 2.876   & 0.644 & &&\\	  	 
\hline
\end{array}
$$
\end{table*}

The most striking characteristic of the CMD of M~22 is the large color
spread of the RGB, incompatible with measurement errors. As summarized
in Section~\ref{intro}, there is a long-standing debate about a
possible metallicity spread in M~22, but the well established presence
of some differential reddening \citep{rich} in the direction of M~22 further
complicates the analysis, since both mechanisms can contribute to widen
the RGB in color and they are difficult to disentangle. 

\begin{figure} 
\includegraphics[width=84mm]{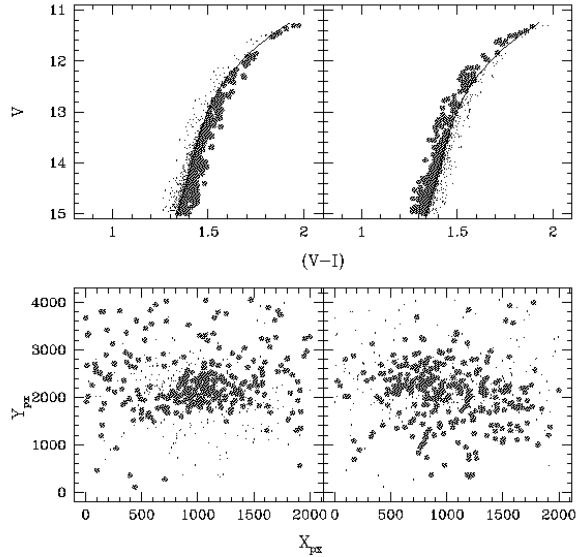}

\caption{The stars in the M~22 RGB are divided into two samples, according to
their position with respect to the mean ridge line (upper panels). The stars 
in the two sample have different spatial distributions (lower panels).} 

\label{dr1} 
\end{figure}

The presence of differential reddening is easily demonstrated in
Fig.~\ref{dr1}. As can be seen, stars on the red and blue sides of the
mean ridge lines derived in Section~\ref{sec-mrl} occupy different
spatial positions in the cluster. In particular, stars {\em redder}
than the mean ridge line tend to populate preferentially the uppermost
part of CCD \#2 (i.e., North of the cluster center). The opposite is
true for stars {\em bluer} than the mean ridge line. Thus, the northern
part of the cluster must be more reddened than the southern part. The
same kind of behaviour can be observed using the HB sequence, which is
less sensitive to metallicity than the RGB, confirming that the
dominant contribution to the color spread of both sequences must be due
to reddening variations. However, we still cannot exclude the presence
of some (small) degree of metallicity spread. 
From a careful inspection of Figure \ref{dr1}, it is also evident that 
differential reddening is present on different scales, ranging from 
$\sim$100 pix (see the {\it ``hole"} at (X, Y) = (600, 1100) in the lower-left
panel) to  $\sim$1000 pix (see the lower half of the lower-left
panel). Therefore, a wide field study is also required in order to obtain
mean properties that are really representative of M~22.

\begin{figure} 
\includegraphics[width=84mm]{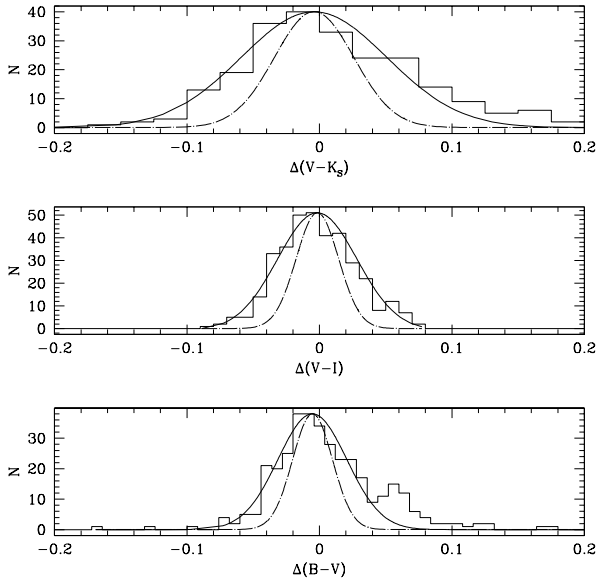}

\caption{The observed color difference distribution between stars in the
RGB and the corresponding mean ridge line is plotted in B-V (lower panel), 
V-I (middle panel) and V-K$_S$ (upper panel). Two gaussian curves
representing the fit to the observed distribution (continuous curve) and the
distribution expected by photometric errors (dot-dashed curve) are also 
plotted on each panel.} 

\label{dr2} 
\end{figure}

Here, we profited from our multi-band photometry to put quantitative
constraints to the amount of metallicity spread allowed by the color
spread of the RGB. In particular, we studied the RGB color distribution
of M~22 in (B$-$V), (V$-$I) and (V$-$K$_S$), selecting stars with V$\leq$15 for
(B$-$V) and (V$-$I), and stars with K$_S$$\leq$12 for (V$-$K$_S$). This is
necessary to avoid any contamination from the galactic bulge and to use
only measurements with the highest possible signal-to-noise ratio. We
then computed the color difference, at fixed magnitude, between each
star and the corresponding mean ridge line (Section~\ref{sec-mrl}). The
histograms of the resulting differences are plotted in Fig.~\ref{dr2}
for each color. A gaussian curve representing the measurement errors is
overplotted on each histogram (dot-dashed curve). A gaussian fit to the
actual color dispersion is also overplotted (solid curve) on each
histogram. The latter gaussian shows of course a larger dispersion than
the one representing the errors only, since it contains also the
contribution of the intrinsic spread, i.e. the differential reddening
plus the eventual metallicity dispersion\footnote{We point out the
presence of a clump of stars with redder colors in Fig.~\ref{dr2},
especially visible in (B$-$V). If the colors of these stars
are due to differential reddening, they could be tracing a denser
interstellar matter region, with an E(B$-$V) which is $\sim$0.06~mag
higher than the average cluster reddening.}. 

We can then represent the observed color dispersion,
$\sigma_{(M_1-M_2)_{obs}}$, as the sum of three terms, one due to
photometric errors, $\sigma_{\delta(M_1-M_2)}$\footnote{Photometric errors are
computed as the standard deviation of repeated measurments of a star
magnitude, available for the B, V, and I filters. In the case of the K$_S$ 
filter, we used errors provided by the 2~MASS extraction algorithm. 
We excluded stars belonging to the inner 1$^{\prime}$ around the cluster center, 
where crowding effects are most severe.}, the second to the
differential reddening, $\sigma_{\Delta E(M_1-M_2)}$, and the third to the
intrinsic metallicity spread, $\sigma_{\Delta\rm{[M/H]}}$ 
\begin{eqnarray*}
\sigma_{(M_1-M_2)_{obs}}^2&=&\sigma_{\delta(M_1-M_2)}^2+\sigma_{\Delta E(M_1-M_2)}^2+\sigma_{\Delta[M/H]}^2\\
&=&\sigma_{\delta(M_1-M_2)}^2+\sigma_{(B-V)_{int}}^2
\end{eqnarray*}
Since the observed color spreads and the photometric errors are
\begin{eqnarray*}
\sigma_{(B-V)_{obs}}=0.026;&~~~&\sigma_{\delta(B-V)}=0.015\\
\sigma_{(V-I)_{obs}}=0.030;&~~~&\sigma_{\delta(V-I)}=0.016\\
\sigma_{(V-K_S)_{obs}}=0.055;&~~~&\sigma_{\delta(V-K_S)}=0.030
\end{eqnarray*}
we can derive the intrinsic color spreads obtaining:
\begin{eqnarray*}
\sigma_{(B-V)_{int}}&=&\sqrt{\sigma_{\Delta E(B-V)}^2+\sigma_{\Delta\rm{[M/H]}}^2}=0.02\\
\sigma_{(V-I)_{int}}&=&\sqrt{\sigma_{\Delta E(V-I)}^2+\sigma_{\Delta\rm{[M/H]}}^2}=0.03\\
\sigma_{(V-K_S)_{int}}&=&\sqrt{\sigma_{\Delta E(V-K_S)}^2+\sigma_{\Delta\rm{[M/H]}}^2}=0.05
\end{eqnarray*}

These values have to be considered as {\em upper limits} to the amount
of differential reddening needed to explain the RGB width, if we assume
zero metallicity spread. The above values are in reasonable agreement
with previous determinations. In fact, $\sigma_{\Delta E(V-I)}$=0.05, derived
by \citet{zoc} using the main sequence, compares reasonably with our
$\sigma_{\Delta E(V-I)}$=0.03, while $\Delta$E(B$-$V)=(0.07;0.08), derived respectively by
\citet{rich} and \citet{twa} from Str\"omgren photometry, compares very well with our
$\Delta$E(B$-$V)$\simeq3\cdot\sigma_{\Delta E(B-V)}$=0.06.

We also note that the intrinsic spread, in the three different colors,
changes following the reddening laws (in the following we will always 
use the reddening laws by \citet{dean} and \citet{sama}). In fact,
assuming $\sigma_{\Delta E(B-V)}$=0.02 as above, we obtain $\sigma_{\Delta E(V-I)}$ and $\sigma_{\Delta E(V-K_S)}$ that
are virtually identical to the values derived above 
\begin{eqnarray*}
\sigma_{\Delta E(V-I)}&=&1.34\cdot\sigma_{\Delta E(B-V)}=0.03\\ 
\sigma_{\Delta E(V-K_S)}&=&2.72\cdot\sigma_{\Delta E(B-V)}=0.06  
\end{eqnarray*}
Therefore, since the intrinsic width of the RGB scales from the optical
planes to the optical-infrared plane exactly as expected from the
reddening laws, we have to conclude that the room left for an intrinsic
metallicity spread must be very small. 

\begin{figure} 
\includegraphics[width=84mm]{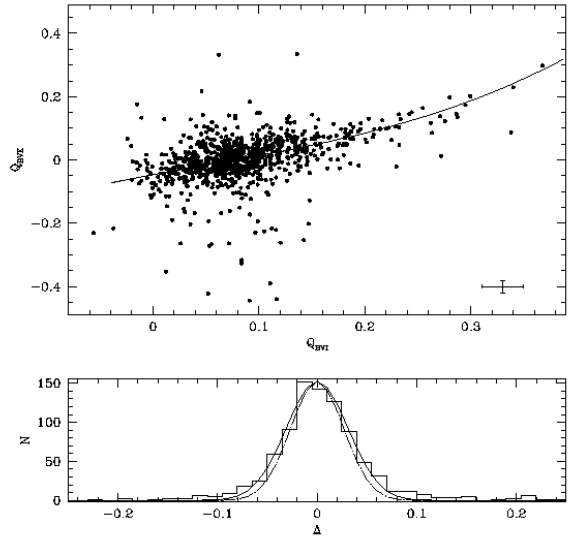}

\caption{RGB stars in M22 are plotted in the plane of the two reddening free 
color indices Q$_{BVI}$ {\it vs} Q$_{BVK}$ (upper panel). The continuos curve 
is the mean locus of stars in the plane. In the lower panel the observed 
distribution of the stars distances from the mean locus is plotted (histogram). 
The two gaussian curves represent a fit to the observed distribution (continuos curve) 
and the distribution expected considering the photometric errors (dotted-dashed
curve).} 

\label{qq} 
\end{figure}

To confirm this conclusion, we made use of two reddening free color
indices
\begin{eqnarray*}
Q_{BVI}&=&(B-V)-\frac{E(B-V)}{E(V-I)}(V-I)\\
Q_{BVK}&=&(B-V)-\frac{E(B-V)}{E(V-K_S)}(V-K_S)
\end{eqnarray*}
and we plotted them in the top panel of Fig.~\ref{qq}. The position of
stars in the $(Q_{BVI},Q_{BVK})$ plane should thus depend only on the
intrinsic properties of the stars (i.e., temperature and chemical
composition). In particular, the spread around the mean locus should
depend only on the eventual metallicity spread and on the photometric
errors, since the Q-color indices are independent on reddening by
definition.

Therefore, we derived the mean ridge line of the locus in Fig.~\ref{qq}
(thick line) with the same method used in Section~\ref{sec-mrl}. We
measured the distance of each star from the mean ridge line and we
constructed the histogram shown in the bottom panel of Fig.~\ref{qq}.
Two gaussians have been overplotted: the one obtained
by propagating the photometric errors to the $(Q_{BVI},Q_{BVK})$ plane
(dotted-dashed line) and the one that best-fits the distribution (solid
line). As can be seen, the two gaussians are virtually identical. 

Summarizing, the metallicity spread contribution to the intrinsic width
of the RGB must be smaller than our typical measurement errors, i.e.,
of the order of $\sim 0.01\div0.02$~mag. 
Using equation 8 in \citet{cb98} and considering (V-I)$_{0,g}$=0.92
(see next section), we derive that a color spread of $\sim 0.01\div0.02$
corresponds to a metallicity spread of approximately
$\Delta$[Fe/H]$\simeq 0.1\div0.2$~dex, which is the maximum metallicity
spread allowed by the present photometry. This spread is also of the order of
the uncertainty of high-resolution abundance determinations for a single star
($\sim$0.15~dex). Therefore such a low metallicity spread, if present, 
could remain hidden into the instrumental errors 
even studying a large sample of high resolution spectra of M~22 stars.  

\section{Metallicity, Reddening and Distance}
\label{sec-par}

In this Section we characterize the photometric properties of the
stellar population in M~22 by measuring the complete set of observables
for the evolved sequences, with the aim of obtaining modern, CCD based
estimates of the cluster mean metallicity, mean reddening and distance
modulus.

The V$_{ZAHB}$ level was obtained 
from the comparison of the observed star distribution
along the HB with synthetic HB of appropriate 
metallicity ($[M/H]\sim-1.5$, see below), following the
semi-empirical approach described by \citet[][]{f99}.   
This
method is extremely robust since it allows the direct determination of
V$_{ZAHB}$, instead of relying on the mean HB level as an indicator of
the ZAHB level. 
We obtained $$V_{ZAHB}=14.33 \pm 0.05$$ about 0.15 mag
fainter than what reported by \citet{harris}. This is easily understood in 
view of the fact that the \citet{harris} value is referred to the mean HB 
level instead of the Zero Age Horizontal branch level \citep[see equation~\#2 in][]{f99}.  

\citet{rosen} derived V$_{ZAHB}=$14.25 for M~22 and, 
considering the zero point difference with our V magnitudes 
(0.05, see section \ref{confronti}), 
our V$_{ZAHB}$ level results to be about 0.03 magnitude fainter than their value. 

Next, we proceeded to the simultaneous determination of the mean
metallicity and reddening, using a procedure that was first introduced
by \citet{s94}. In its original description, the method allowed the
derivation of [Fe/H] in the \citet{zw} scale, using V,(V$-$I)
photometries and mean ridge lines of the RGB. Later, the procedure was
calibrated to the \citet{cg97} scale by \citet{cb98} and 
extended to the B,(B$-$V) plane by \citet{sala} and \citet{f99}.

Using the calibration of \citet{cb98}, we assumed a mean HB level
$<\rm{V}_{HB}>=$14.15 \citep{harris} and used the V,(V$-$I) mean ridge line
derived in Section~\ref{sec-mrl}, obtaining $\rm{[Fe/H]}_{CG}=-1.63$ and
$E(B-V)=0.36$ (we also derived  (V-I)$_{0,g}$=0.92 and
$\Delta$V$_{1.2}$=2.19)\footnote{(V-I)$_{0,g}$ and (B-V)$_{0,g}$ are the 
dereddened RGB colors at the level of the HB, while  $\Delta$V$_{1.2}$ is the
difference in V between the HB and the RGB  at a dereddened color 1.2
\citep[i.e., (V-I)$_0$=1.2 and (B-V)$_0$=1.2, see][]{s94,f99}.}.  Using the
calibration by \citet{f99}, we adopted the V$_{ZAHB}$ just derived and used the
V,(B$-$V) mean ridge line, obtaining $\rm{[Fe/H]}_{CG}=-1.73$ and $E(B-V)=0.39$
(we also derived (B-V)$_{0,g}$=0.73 and $\Delta$V$_{1.2}$=2.68)

Since the two above reddening determinations are in good agreement with
each other, we will adopt their average for the rest of this article:
$$E(B-V)=0.38\pm 0.02$$

Our determination of the reddening is identical to the one by \citet[][]{rich}
which  is also the most recent determination. Other estimates of the  reddening
range from E(B$-$V)=0.32 to 0.42 and can be found in 
\citet{hesser2,racine,harris} and \citet{crok}.   These values are in broad
agreement with our determination within the errors, if we  take into account  
the presence of a strong differential reddening in direction of M~22 (see
previous section).

\begin{figure} 
\includegraphics[width=84mm]{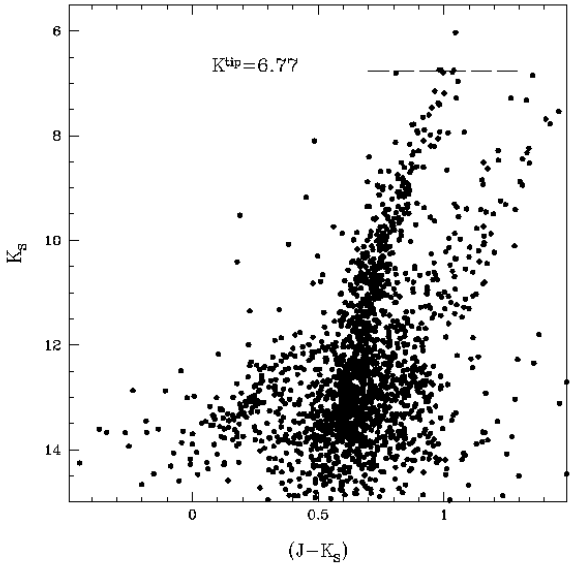}

\caption{Infrared M22 CMD. The dashed line indicate the putative position of the
RGB-Tip.}  

\label{tipK} 
\end{figure}

The average metallicity of M~22 can also be derived from the
optical-infrared CMDs V,(V$-$K$_S$) and K$_S$,(J$-$K$_S$) using the RGB intrinsic
colors (V$-$K$_S$)$_{0}$ and (J$-$K$_S$)$_{0}$, measured at different magnitude
levels \citep[see the definition of these parameters in][]{f00}. 
We used the calibrations of \citet{elena} to yield metallicities 
in the \citet{cg97} scale from the infrared photometry in the 2~MASS photometric
system. We corrected our optical-infrared CMD using 
$E(B-V)=0.38$ and the (m$-$M)$_0$ derived below. This way we obtained different photometric metallicity estimates for M~22,
which produce an average value of: $\rm{[Fe/H]}_{CG}=-1.68$. 

Averaging together all the optical and infrared metallicity determinations we obtained: 
$$\rm{[Fe/H]}_{CG}=-1.68 \pm 0.15$$ 
This value is in agreement, within the errors, with the most recent spectroscopic and 
photometric works which provide mean metallicities of -1.55, -1.48
\citep[][respectively]{leh,cg97} and -1.62 \citep{rich}. 

\begin{figure} 
\includegraphics[width=84mm]{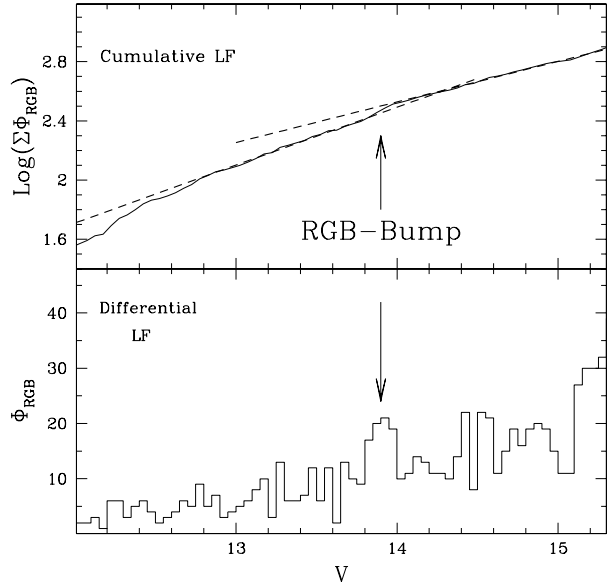}

\caption{Differential (lower panel) and cumulative (upper panel)
luminosity  functions for stars selected on the RGB
of M22.  The arrows indicate the location of the RGB-bump at
V=13.90$\pm$0.05.} 

\label{bumpV} 
\end{figure} 

An estimate of the global metallicity 
can be now derived according to the prescription of \citet{scs93}:
$$\rm{[M/H]}=\rm{[Fe/H]}+\log(0.638\cdot10^{\rm{[\alpha/Fe]}}+0.362)=-1.47$$ 
by assuming   [$\alpha$/Fe]=+0.30 \citep{sc96}.
The distance modulus is   derived  from the comparison
of the observed value of the ZAHB obtained above 
($V_{ZAHB}=14.33 \pm 0.05$)
and the 
absolute level computed from the \citet[][]{SCL} models 
\citep[see Eq 4 by ][]{f99}.
From this relation and the global metallicity
computed above ($\rm{[M/H]}=-1.47$) we  get
$M_V^{ZAHB}=0.59$. From this figure we 
derived an apparent  distance modulus 
 $$(m-M)_V=13.74\pm 0.2$$ and finally 
 (using E(B$-$V)=0.38), an intrinsic distance
modulus (m-M)$_0$=12.56 $\pm$ 0.2 which corresponds to a
distance of $\sim 3.2$ Kpc.
A conservative uncertainty of 0.2 mag is assumed. Our $(m-M)_V$
is about 0.15 mag fainter than the value reported by \citet{harris}. 
Therefore the two distance moduli are in agreement within the uncertainties, however 
a systematic difference between the \citet{harris} distance scale 
and distances derived by our semi-empirical procedure still remains and is 
further discussed in \citet{f99}. 

As a consistency check for the parameters derived until now, we derived
the putative position of the RGB-tip, in the K band. From the relations
provided by \citet{f00} and assuming the [Fe/H]$_{CG}$ just derived, we 
obtained as absolute magnitude of the RGB-tip: $$M_{K}^{tip}=-5.93$$ 
Then, using the (m-M)$_0$ and E(B$-$V) we derived the predicted position for the
RGB-Tip in apparent magnitude: $$K^{tip}=6.77$$

In Fig.~\ref{tipK} the K$_S$,(J$-$K$_S$) CMD of
M~22 is plotted. Even if the presented CMD does not possess enough stars
in the upper RGB for a safe determination of the RGB-Tip, it is
nevertheless clear that the predicted RGB-tip is very similar to the
magnitude at which the star counts drop to zero, K$_S\simeq$6.74.

The last important feature that we were able to measure on the RGB of
M~22 is the so-called RGB-bump. This feature was one of the first
successful predictions of the stellar evolution theory
\citep{thomas,iben}, identified observationally in 47~Tuc and
subsequently in all the properly observed clusters
\citep{kingbump,fp90,f99,zoc99,f00,bumpk}, both in the optical and infrared
filters. Recently, the RGB-bump was also observed in a few satellites
of the  Milky Way \citep[][ and references therein]{io,draco}.

\begin{figure} 
\includegraphics[width=84mm]{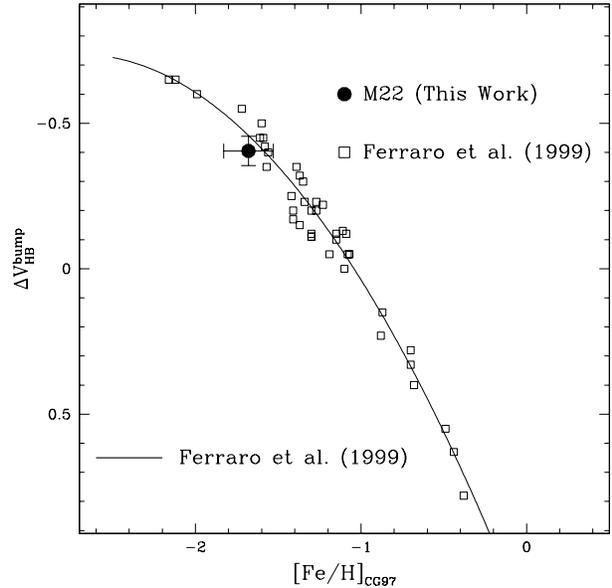}

\caption{The magnitude of the M~22 RGB-bump (large filled circle) is compared 
with that of other globular clusters and with the empirical
calibration of \citet{f99} in the 
$\Delta V^{bump}_{HB}$ {\it vs} [Fe/H]$_{CG97}$ plane.} 

\label{bump2} 
\end{figure} 

As demonstrated by \citet{fp90}, the change in slope of the integrated
luminosity function is the safest way to identify the RGB-bump
location, since it makes use of stars contained in several magnitude
bins. We thus identified the RGB-bump of M~22 in the V\ band, as shown
in Fig.~\ref{bumpV}:
$$
V^{bump}=13.90\pm 0.05
$$
We also identified the RGB-bump in the K$_S$ band, finding a value in good 
agreement with \citet[][]{bumpk}.

In Figure~\ref{bump2} the
difference of V$^{bump}$ and V$_{ZAHB}$, $\Delta$V$^{bump}_{HB}$ versus
[Fe/H] in the \citet{cg97} scale, is plotted for the 42 clusters of 
\citet[][ small empty squares]{f99} and for M~22 (big filled circle). 
As can be seen, M~22 matches quite well the empirical calibration of
\citet{f99}.


\begin{figure} 
\includegraphics[width=84mm]{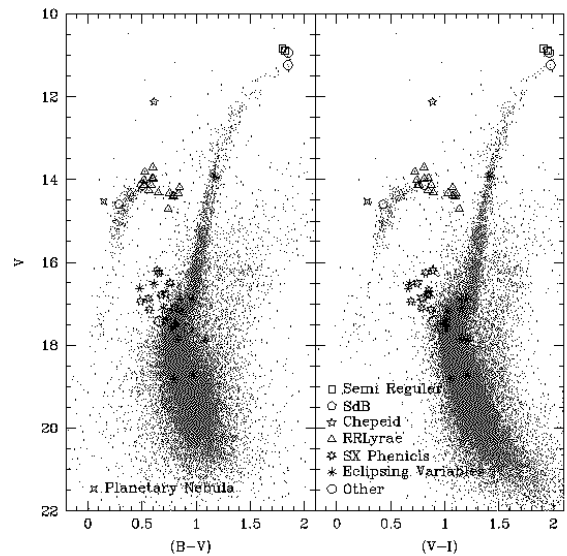}

\caption{The position of the variable stars recovered in our catalogue is
plotted in the optical CMDs. Different classes of
variables have been plotted using different symbols.} 

\label{var} 
\end{figure} 

\section{Variable Stars and Peculiar Objects}
\label{sec-var}

M~22 is known to host various type of variable stars, from the
RR~Lyrae stars typical of globular clusters \citep{rr} to rather peculiar or rare objects 
such as a type II cepheid, one sdB star \citep{rr,kalu} and a Planetary Nebula 
(see next section). A probable dwarf nova in outburst phase has also been recently
identified \citep{dn} as well as a population of X-ray sources \citep{webb}. 
In spite of the modest number of Blue Straggler stars, 
M~22 contains also a significant number of SX Phoenicis variables \citep{kalu}.  

We cross-correlated our catalogue with that of variable stars
by \citet{clem} which provides coordinates, classification and other useful 
information for all the known variable stars in M~22.  
In Fig. \ref{var} we used various symbols to show the position in the CMD of 
the various type of variables successfully identified in our
photometry. 
51 variables over the 79 known in M~22 have been identified. Among them, 16 are 
RR~Lyrae stars. Averaging the V magnitudes of the RR~Lyrae observed at random phase in our 
photometry we obtain V$_{HB}$=14.17$\pm$0.25. This value for the HB magnitude is about 
0.15 magnitude brighter than our estimate of V$_{ZAHB}$ in qualitative agreement to what expected 
from equation \#2 by \citet[][]{f99}.


\begin{figure} 
\includegraphics[width=84mm]{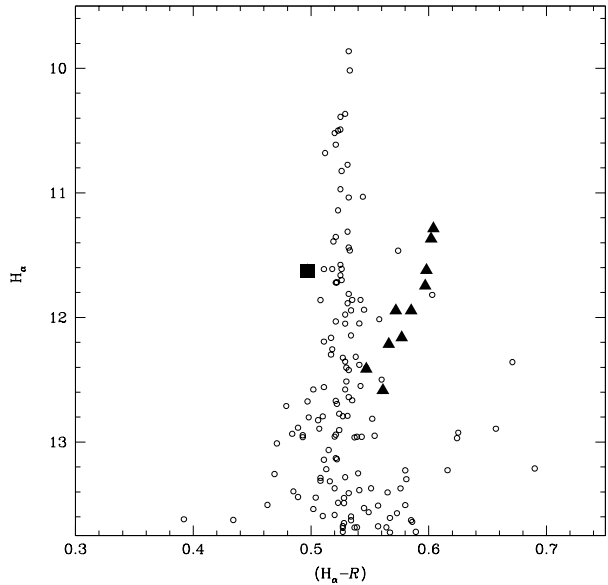}
\caption{(H$_{\alpha}$; H$_{\alpha}-\Re$) CMD for M~22 stars lying within 150 pixels
from S$_\star$ (small empty circles). S$_\star$, the optical counterpart of the PN 
in M~22, is represented as a large filled square, while stars in the HB phase are 
plotted as filled triangles.} 

\label{ha} 
\end{figure} 

\subsection{The Planetary Nebula in M22} 
\label{sec-pn}

Together with Pal~6, NGC~6441 and M~15 \citep{pal,m15}, 
M22 is one of the few galactic globular clusters  which are known to host
a planetary nebula (PN). It was first discovered with the IRAS
satellite by \citet{G86} as a pointlike source (IRAS~18333-2357) and
then identified as a Planetary Nebula \citep{G89}. The central star of
the PN has also been identified \citep{G89}, as a blue star (the {\em
``southern component''}, hereafter S$_\star$) belonging to a pair  which lies
only $\sim 2\arcsec$ away from the infrared source. 

Some of the physical properties of the PN and its central star were
derived by \citet{cohen89}. Considering the expansion velocity 
(11~km/s) of the nebula and assuming that its
size corresponds to the distance to S$_\star$, \citet{cohen89} 
concluded that the
age of the PN appears to be only $\simeq$ 6000 yr. This short time
scale implies that the central star should still be quite bright, with
a luminosity comparable to that of the RGB-tip. This poses an apparent
problem, since S$_\star$ has a magnitude (and even colors) comparable
to that of a slightly evolved HB star, as can be seen in
Fig.~\ref{var}. 

To further investigate the photometric properties of S$_\star$, we
constructed the instrumental (non-calibrated)
H$_{\alpha}$,(H$_{\alpha}-\Re$) CMD for stars contained in a circle
with a radius of 150~pixels, centered on S$_\star$ (Fig.~\ref{ha}). In
this diagram, RGB stars occupy a vertical sequence while genuine HB
stars (filled triangles), which have a more pronounced H$_{\alpha}$ absorption, tend to
have redder (H$_{\alpha}-\Re$) colors. In the same diagram, S$_\star$ 
(large filled square) 
shows a moderate H$_{\alpha}$ excess, with (H$_{\alpha}-\Re$) slightly
bluer than the normal RGB stars and very different from the HB stars.
This is in agreement with the substantial amount of hydrogen emission displayed
by the spectrum of S$_\star$ \citep{har93}, consistent with what
expected if S$_\star$ is the central star of a new-born PN. 

\begin{figure}  \includegraphics[width=84mm]{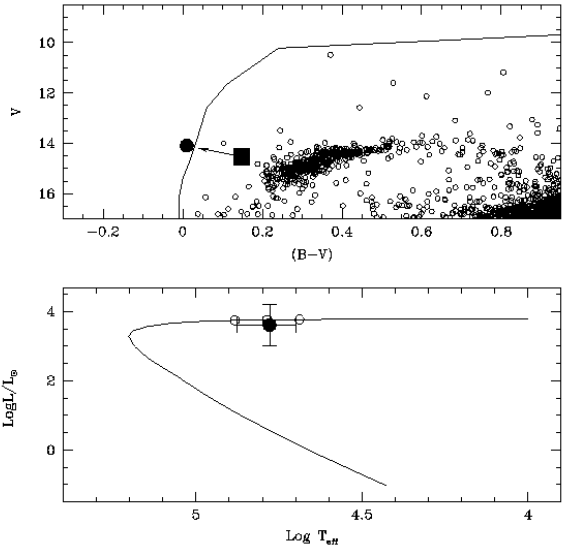}

\caption{Upper panel: zoomed CMD in the HB region. 
The position of S$_\star$ is marked as a filled square. The filled circle 
represents the putative position of S$_\star$ if a reddening excess of 
$\Delta$E(B-V)$\simeq$0.14 is assumed. 
The continuous curve is a post-AGB theoretical isochrone 
with t$\simeq$12.5~Gyr and Z=0.0004. Lower panel: 
The position of S$_\star$ in the absolute plane, once applied the 
bolometric correction, is plotted (filled circle). The continuous curve is 
a theoretical post-AGB evolutionary track for a 1~M$_{\odot}$ and Z=0.001 star.
Empy circles mark evolutionary steps between 5000 and 8300 years.} 

\label{bol} 
\end{figure} 

We thus need to explain the unusual position of S$_\star$ in the
V,(B$-$V) diagram. An enlargement of the CMD around S$_\star$ is shown
in Fig.~\ref{bol} (upper panel), where the theoretical isochrone of a
post-AGB star with t$\simeq$12.5~Gyr and Z=0.0004 has been also
overplotted \citep{iso}. Clearly, the position of S$_\star$ (large filled 
square) is not compatible with the isochrone. However, S$_\star$ should
be surrounded by nebular dust, which causes an excess of reddening of
about $\Delta$E(B-V)$\simeq$0.14 \citep{cud90,har93}. After applying
the corresponding correction, the position of S$_\star$ is reconciled
with the isochrone (large filled circle in Fig.~\ref{bol}, upper
panel). 

The theoretical ($\log T_{eff}$,$\log L/L_{\odot}$) plane can also provide 
useful insight into the evolutionary state of S$_\star$ and its compatibility 
with a post AGB-star of about 6000yr. 
Spectroscopic studies derive for
S$_\star$ a temperature of 50,000$\leq$T$_{eff}\leq$75,000 K
\citep{cohen89,har93}. In
this range of temperatures, the bolometric correction in the V band,
BC$_V$, varies between -4.08 and -5.30, using a blackbody model \citep{livia}. 
If we assume  T$_{eff}$=60,000~K (hence BC$_V$=-4.62),
$\Delta$E(B-V)$\simeq$0.14 and (m-M)$_V$=13.75, then S$_\star$ occupies
the position marked by a large filled circle in the lower panel of
Fig.~\ref{bol}. We also overplotted the theoretical post-AGB
evolutionary track for a 1~M$_{\odot}$ and Z=0.001 star
\citep{pntrack}. Tiny empty dots mark the positions occupied by stars
of 5500, 6900 and 8300~yr. As can be seen, S$_\star$ closely matches
the track at a luminosity comparable to that of the RGB-tip and a
temperature which is fully compatible with the central star of a PN of
$\sim$6000~yr. 

We thus have been able to add supporting evidence to the identification
of S$_\star$ as the central star of the planetary nebula
IRAS~18333-2357 in M~22.


\section{Summary and Conclusions}  
\label{sec-con}

We presented a wide field (33$\arcmin\times$34$\arcmin$), multi-band photometry 
(B, V, I, H$_{\alpha}$ and the 
adjacent continuum, $\Re$) of the globular
cluster M~22. For the H$_{\alpha}$ and $\Re$ filters we presented only 
instrumental magnitudes in a tiny area around the PN in M~22. 
We provided the astrometrically calibrated catalog containing B, V and
I calibrated magnitudes for $\sim$140,000 stars covering an area of 
$\sim$24$\arcmin\times$33$\arcmin$. 
About 2000 stars in our catalog have
been also measured in the near infrared by the 2~MASS survey, providing
calibrated J, H and K$_S$ magnitudes in addition to our five optical
bands. 

We use this catalog to characterize the evolved stellar sequences in
M~22, especially the RGB, by deriving mean ridge lines in the optical
and infrared colors, and by measuring the V magnitude of the
RGB-bump, which appears in good agreement with the most recent
calibration versus metallicity \citep{f99}. We also derived 
the mean metallicity, reddening and distance
moduli, in a self-consistent way. The derived values agree well with 
previous determinations.

Profiting from our multi-band catalogue, we re-examined the problem of
the metallicity spread in M~22. We demonstrated the presence of
differential reddening, confirming earlier findings \citep[][ and
references therein]{rich}. We also conclude that, according to our
photometric measurements, most of the intrinsic width of the RGB must
be due to differential reddening, while the maximum metallicity spread
allowed by our data is $\Delta$[Fe/H]$\simeq 0.1\div 0.2$~dex, i.e.,
compatible with the photometric errors.

We finally identified most of the variable
stars and peculiar objects known in the observed field of M~22. In
particular, we provided additional evidence supporting the optical
identification of the central star in the planetary nebula
IRAS~18333-2357, one of the few 
identified in a GGC up to now.


\section*{Acknowledgments}

We are grateful to L.~Origlia and E.~Valenti for useful suggestions and discussions.
We also thank an anonymous referee for useful comments that have substantially
improved our paper. 
This research is partially supported by MIUR (Ministero della 
Istruzione, dell'Universit\`a e della Ricerca) and ASI (Agenzia 
Spaziale Italiana). Part of the data analysis has been performed using 
software packages developed by P. Montegriffo at the Osservatorio 
Astronomico di Bologna.




\label{lastpage}


\begin{thebibliography}{99}

\bibitem[\protect\citeauthoryear{Adams et al.}{1984}]{m15} Adams, S., Seaton, M.~J., 
Howarth, I.~D., Auriere, M., \& Walsh, J.~R.\ 1984, MNRAS, 207, 471 

\bibitem[\protect\citeauthoryear{Albrow, De Marchi, \& Sahu}{2002}]{dina} Albrow, 
M.~D., De Marchi, G., \& Sahu, K.~C.\ 2002, ApJ, 579, 660 

\bibitem[\protect\citeauthoryear{Anderson et al.}{2003}]{dn} Anderson, J., Cool, A.~M., \& 
King,  I.~R.\ 2003, ApJ, 597, L140 

\bibitem[\protect\citeauthoryear{Anthony-Twarog, Twarog, \& Craig}{1995}]{twa} 
Anthony-Twarog, B.~J., Twarog, B.~A., \& Craig, J.\ 1995, PASP, 107, 32 

\bibitem[\protect\citeauthoryear{Arp \& Melbourne}{1959}]{arp} Arp, H.~C.~\&  Melbourne, W.~G.\
1959, AJ, 64, 28 

\bibitem[\protect\citeauthoryear{Bellazzini et al.}{2002}]{draco} Bellazzini, M.,  Ferraro,
F.~R., Origlia, L., Pancino, E., Monaco, L., \& Oliva, E.\ 2002,  AJ,
124, 3222 

\bibitem[\protect\citeauthoryear{Bertelli et al.}{1994}]{iso} Bertelli, G., Bressan,  A.,
Chiosi, C., Fagotto, F., \& Nasi, E.\ 1994, A\&AS, 106, 275 

\bibitem[\protect\citeauthoryear{Brown \& Wallerstein}{1992}]{brown} Brown, J.~A.~\& 
Wallerstein, G.\ 1992, AJ, 104, 1818 

\bibitem[\protect\citeauthoryear{Cannon}{1980}]{cannon80} Cannon, R.~D.\ 1980, A\&A, 81, 379 

\bibitem[\protect\citeauthoryear{Carpenter}{2001}]{carp} Carpenter, J.~M.\ 2001, AJ, 121, 2851 

\bibitem[\protect\citeauthoryear{Carretta \& Bragaglia}{1998}]{cb98} Carretta, E.~\&  Bragaglia,
A.\ 1998, A\&A, 329, 937 

\bibitem[\protect\citeauthoryear{Carretta \& Gratton}{1997}]{cg97} Carretta, E.~\&  Gratton,
R.~G.\ 1997, A\&AS, 121, 95 

\bibitem[\protect\citeauthoryear{Cho \& Lee}{2002}]{bumpk} Cho, D.~\& Lee, S.\ 2002,  AJ, 124,
977 

\bibitem[\protect\citeauthoryear{Clement et al.}{2001}]{clem} Clement, C.~M.~et al.\  2001, AJ,
122, 2587 

\bibitem[\protect\citeauthoryear{Cohen}{1981}]{cohen} Cohen, J.~G.\ 1981, ApJ, 247,  869 

\bibitem[\protect\citeauthoryear{Cohen \& Gillett}{1989}]{cohen89} Cohen, J.~G.~\&  Gillett,
F.~C.\ 1989, ApJ, 346, 803 

\bibitem[\protect\citeauthoryear{Crocker}{1988}]{crok} Crocker, D.~A.\ 1988, AJ, 96, 
1649 

\bibitem[\protect\citeauthoryear{Cudworth}{1990}]{cud90} Cudworth, K.~M.\ 1990, AJ,  99, 1863 

\bibitem[\protect\citeauthoryear{Da Costa \& Armandroff}{1990}]{m2} Da Costa,  G.~S.~\&
Armandroff, T.~E.\ 1990, AJ, 100, 162 

\bibitem[\protect\citeauthoryear{Davidge \& Harris}{1996}]{davi1} Davidge, T.~J.~\& 
Harris, W.~E.\ 1996, ApJ, 462, 255 

\bibitem[\protect\citeauthoryear{Davidge \& Harris}{1995}]{davi2} Davidge, T.~J.~\& 
Harris, W.~E.\ 1995, ApJ, 445, 211 

\bibitem[\protect\citeauthoryear{Dean, Warren \& Cousins}{1978}]{dean} Dean, J.F., Warren,
P.R.,  \& Cousins, A.W., 1978, MNRAS, 183, 569 

\bibitem[\protect\citeauthoryear{Dickens \& Woolley}{1967}]{dickens} Dickens, R.~J.~\&  Woolley,
R.~v.~d.~R.\ 1967, Royal Greenwich Observatory Bulletin, 128, 255 

\bibitem[\protect\citeauthoryear{Elias et al.}{1982}]{eli} Elias, J.~H., Frogel, J.~A., 
Matthews, K., \& Neugebauer, G.\ 1982, AJ, 87, 1029 

\bibitem[\protect\citeauthoryear{Ferraro et al.}{1992}]{f2}  Ferraro, F.~R., Fusi Pecci, F., \&
Buonanno, R.\ 1992, MNRAS, 256, 376 

\bibitem[\protect\citeauthoryear{Ferraro et al.}{1997}]{f1} Ferraro, F.~R.,  Carretta, E.,
Corsi, C.~E., Fusi Pecci, F., Cacciari, C., Buonanno, R.,  Paltrinieri,
B., \& Hamilton, D.\ 1997, A\&A, 320, 757 

\bibitem[\protect\citeauthoryear{Ferraro et al.}{1999}]{f99} Ferraro, F.~R., Messineo, M.,  Fusi
Pecci, F., de Palo, M.~A., Straniero, O., Chieffi, A., Limongi, M. 

\bibitem[\protect\citeauthoryear{Ferraro et al.}{2000}]{f00} Ferraro, F.R., Montegriffo, P., 
Origlia, L., \& Fusi Pecci, F., 2000, AJ, 119, 1282 

\bibitem[\protect\citeauthoryear{Ferraro et al.}{2001}]{f01} Ferraro, F.~R.,  D'Amico, N.,
Possenti, A., Mignani, R.~P., \& Paltrinieri, B.\ 2001, ApJ,  561,
337 

\bibitem[\protect\citeauthoryear{Fusi Pecci et al.}{1990}]{fp90} Fusi Pecci, F., Ferraro,
F.~R.,  Crocker, D.~A., Rood, R.~T., Buonanno, R.\ 1990, A\&A, 238, 95 

\bibitem[\protect\citeauthoryear{Gillett et al.}{1986}]{G86} Gillett, F.~C., Backman, D.~E., 
Beichman, C., \& Neugebauer, G.\ 1986, ApJ, 310, 842 

\bibitem[\protect\citeauthoryear{Gillett et al.}{1989}]{G89} Gillett, F.~C., Jacoby,  G.~H.,
Joyce, R.~R., Cohen, J.~G., Neugebauer, G., Soifer, B.~T., Nakajima, 
T., \& Matthews, K.\ 1989, ApJ, 338, 862 

\bibitem[\protect\citeauthoryear{Gratton}{1982}]{gratton} Gratton, R.~G.\ 1982, A\&A, 115, 171 

\bibitem[\protect\citeauthoryear{Gratton \& Ortolani}{1989}]{go89} Gratton, R.~G.~\&  Ortolani,
S.\ 1989, A\&A, 211, 41 

\bibitem[\protect\citeauthoryear{Harrington \& Paltoglou}{1993}]{har93} Harrington,  J.~P.~\&
Paltoglou, G.\ 1993, ApJ, 411, L103 

\bibitem[\protect\citeauthoryear{Harris}{1996}]{harris} Harris, W.~E.\ 1996, VizieR  Online Data
Catalog, 7195 

\bibitem[\protect\citeauthoryear{Harris \& Racine}{1979}]{racine} Harris, W.~E.~\& 
Racine, R.\ 1979, ARA\&A, 17, 241 

\bibitem[\protect\citeauthoryear{Hesser et al.}{1977}]{hesser} Hesser,  J.~E., Hartwick,
F.~D.~A., \& McClure, R.~D.\ 1977, ApJS, 33, 471 

\bibitem[\protect\citeauthoryear{Hesser}{1976}]{hesser2} Hesser, J.~E.\ 1976, PASP, 88, 
849 

\bibitem[\protect\citeauthoryear{Iben}{1968}]{iben} Iben, I.~Jr.\ 1968, Nature, 220, 143 

\bibitem[\protect\citeauthoryear{Jacoby et al.}{1997}]{pal} Jacoby, G.~H., Morse, 
J.~A., Fullton, L.~K., Kwitter, K.~B., \& Henry, R.~B.~C.\ 1997, AJ, 114, 
2611 

\bibitem[\protect\citeauthoryear{Kaluzny \& Thompson}{2001}]{kalu} Kaluzny, J.~\&  Thompson,
I.~B.\ 2001, A\&A, 373, 899 

\bibitem[\protect\citeauthoryear{King et al.}{1985}]{kingbump} King,  C.~R., Da Costa, G.~S., \&
Demarque, P.\ 1985, ApJ, 299, 674 

\bibitem[\protect\citeauthoryear{Laird, Wilhelm, \& Peterson}{1991}]{laird} Laird, J.~B.,
Wilhelm, R.~J., \& Peterson, R.~C.\ 1991, ASP Conf.~Ser.~ 13: The 
Formation and Evolution of Star Clusters, 578 

\bibitem[\protect\citeauthoryear{Landolt}{1992}]{landolt} Landolt, A.~U.\ 1992, AJ,  104, 340 

\bibitem[\protect\citeauthoryear{Lee \& Carney}{1999}]{m2bv} Lee, J.~\& Carney, 
B.~W.\ 1999, AJ, 117, 2868 


\bibitem[\protect\citeauthoryear{Lehnert, Bell, \& Cohen}{1991}]{leh} Lehnert, M.~D., Bell,
R.~A., \& Cohen, J.~G.\ 1991, ApJ, 367, 514 

\bibitem[\protect\citeauthoryear{Manduca \& Bell}{1978}]{manduca} Manduca, A.~\& Bell,  R.~A.\
1978, ApJ, 225, 908 

\bibitem[\protect\citeauthoryear{Monaco et al.}{2002}]{io} Monaco, L., Ferraro, F.~R.,
Bellazzini,  M., \& Pancino, E.\ 2002, ApJ, 578, L47 

\bibitem[\protect\citeauthoryear{Norris \& Freeman}{1983}]{nofre83} Norris, J.~\& Freeman,
K.~C.\ 1983, ApJ, 266, 130 

\bibitem[\protect\citeauthoryear{Origlia \& Leitherer}{2000}]{livia} Origlia, L.~\& 
Leitherer, C.\ 2000, AJ, 119, 2018 

\bibitem[\protect\citeauthoryear{Peterson}{1980}]{ruth} Peterson, R.~C.\ 1980, IAU  Symp.~ 85:
Star Formation, 85, 461 

\bibitem[\protect\citeauthoryear{Pilachowski, Leep, Wallerstein, \& Peterson}{1982}]{cathy}
Pilachowski, C., Leep, E.~M.,  Wallerstein, G., \& Peterson, R.~C.\
1982, ApJ, 263, 187 

\bibitem[\protect\citeauthoryear{Piotto \& Zoccali}{1999}]{zoc} Piotto, G.~\&  Zoccali, M.\
1999, A\&A, 345, 485 

\bibitem[\protect\citeauthoryear{Richter, Hilker, \& Richtler}{1999}]{rich} Richter,  P.,
Hilker, M., \& Richtler, T.\ 1999, A\&A, 350, 476 

\bibitem[\protect\citeauthoryear{Rosenberg et al.}{2000}]{rosen}  Rosenberg, A., Piotto, G.,
Saviane,  I., \& Aparicio, A.\ 2000, A\&AS, 144, 5 

\bibitem[\protect\citeauthoryear{Sahu et al.}{2001}]{sahu} Sahu, K.~C., Casertano, 
S., Livio, M., Gilliland, R.~L., Panagia, N., Albrow, M.~D., \& Potter, M.\ 
2001, Nature, 411, 1022 

\bibitem[\protect\citeauthoryear{Salaris \& Cassisi}{1996}]{sc96} Salaris, M.~\&  Cassisi, S.\
1996, A\&A, 305, 858 

\bibitem[\protect\citeauthoryear{Salaris, Chieffi, \& Straniero}{1993}]{scs93}  Salaris, M.,
Chieffi, A., \& Straniero, O.\ 1993, ApJ, 414, 580 

\bibitem[\protect\citeauthoryear{Sarajedini}{1994}]{s94} Sarajedini, A.\ 1994, AJ,  107, 618 

\bibitem[\protect\citeauthoryear{Sarajedini \& Layden}{1997}]{sala} Sarajedini, A.~\& 
Layden, A.\ 1997, AJ, 113, 264 

\bibitem[\protect\citeauthoryear{Savage \& Mathis}{1979}]{sama} Savage, B.~D.~\&  Mathis, J.~S.\
1979, ARA\&A, 17, 73 

\bibitem[\protect\citeauthoryear{Shapley}{1930}]{shapley} Shapley, H.\ 1930, Harvard College
Observatory Bulletin, 874, 4 

\bibitem[\protect\citeauthoryear{Stetson}{1987}]{daophot} Stetson, P.~B.\ 1987, PASP,  99, 191
\ 1999, AJ, 118, 1738 

\bibitem[\protect\citeauthoryear{Straniero, Chieffi, \& Limongi}{1997}]{SCL} 
Straniero, O., Chieffi, A., \& Limongi, M.\ 1997, ApJ, 490, 425 

\bibitem[\protect\citeauthoryear{Thomas}{1967}]{thomas} Thomas, H.-C. \ 1967, Z. Astrophys, 67,
420 

\bibitem[\protect\citeauthoryear{Valenti et al.}{2004, in preparation}]{elena} Valenti, E., Ferraro, F.~R. 
and Origlia, L.\ 2004, in preparation

\bibitem[\protect\citeauthoryear{Vassiliadis \& Wood}{1994}]{pntrack} Vassiliadis, E.~\&  Wood,
P.~R.\ 1994, VizieR Online Data Catalog, 209, 20125

\bibitem[\protect\citeauthoryear{Webb, Gendre, \& Barret}{2002}]{webb} Webb, N.~A., 
Gendre, B., \& Barret, D.\ 2002, A\&A, 381, 481 

\bibitem[\protect\citeauthoryear{Wehlau \& Hogg}{1978}]{rr} Wehlau, A.~\& Hogg, 
H.~S.\ 1978, AJ, 83, 946 

\bibitem[\protect\citeauthoryear{Woolley}{1966}]{woolley} Woolley, R.~R.\ 1966, Royal 
Observatory Annals, 2, 1 

\bibitem[\protect\citeauthoryear{Zinn \& West}{1984}]{zw} Zinn, R.~\& West, M.~J.\  1984, ApJS,
55, 45 

\bibitem[\protect\citeauthoryear{Zoccali et al.}{1999}]{zoc99} Zoccali, M., Cassisi,  S.,
Piotto, G., Bono, G., \& Salaris, M.\ 1999, ApJ, 518, L49 


\end{thebibliography}
\end{document}